\documentclass[12pt,draftclsnofoot,onecolumn,notitlepage]{IEEEtran}

\IEEEoverridecommandlockouts

\usepackage{cite}
\usepackage{ieeefig}
\usepackage{amsfonts}
\usepackage{epsfig,color}
\usepackage{graphicx}
\usepackage[caption=false,font=footnotesize]{subfig}
\usepackage{stfloats}
\usepackage{amsmath}
\usepackage{paralist}
\usepackage{color}

\newtheorem{theorem}{\bf Theorem}

\newtheorem{remark}{\bf Remark}

\begin{document}
%
\title{\LARGE{The degrees of freedom of MIMO networks with full-duplex receiver cooperation but no CSIT} }

\author{Chinmay S.~Vaze %
        and~Mahesh K.~Varanasi
\thanks{
This work was presented at the Workshop on Interference in Wireless Networks, Boston University, June 2012.

Chinmay S. Vaze was with the Department of Electrical, Computer, and Energy
Engineering, University of Colorado, Boulder, CO 80309-0425 USA. He is now with Qualcomm Inc., San Diego, CA. (e-mail: Chinmay.Vaze@colorado.edu). }

\thanks{
Mahesh K. Varanasi is with the Department of Electrical, Computer, and Energy Engineering, University of Colorado, Boulder, CO 80309-0425 USA (e-mail: varanasi@colorado.edu).}}

%



\maketitle

\begin{abstract}
The question of whether the degrees of freedom (DoF) of multi-user networks can be enhanced even under isotropic fading and no channel state information (or output feedback) at the transmitters (CSIT) is investigated. Toward this end, the two-user MIMO (multiple-input, multiple-output) broadcast and interference channels are studied with no side-information at the transmitters and with receivers equipped with full-duplex radios. The full-duplex feature allows for receiver cooperation because each receiver, in addition to receiving the signals sent by the transmitters, can also simultaneously transmit a signal in the same band to the other receiver. Unlike the case of MIMO networks with CSIT and full-duplex receivers, for which DoF are known, it is shown that for MIMO networks with no CSIT, full-duplex receiver cooperation is beneficial to such an extent that even the DoF region is enhanced. Indeed, for important classes of two-user MIMO broadcast and interference channels, defined by certain relationships on numbers of antennas at different terminals, the exact DoF regions are established. The key to achieving DoF-optimal performance for such networks are new retro-cooperative interference alignment schemes. Their optimality is established via the DoF analysis of certain genie-aided or enhanced version of those networks.
\end{abstract}

\begin{IEEEkeywords}
Broadcast channel, Channel state information, Degrees of freedom, Full-duplex, Interference alignment, Interference channel, MIMO, Receiver cooperation.
\end{IEEEkeywords}


\newpage

\section{Introduction}


\IEEEPARstart{T}{he} degrees of freedom (DoF) regions of the MIMO broadcast channel (BC) and the MIMO interference channel (IC) are known when perfect and instantaneous channel state information (CSI) is available at all terminals of the network \cite{Shamai-W-S, Chiachi-Jafar, Jafar-Maralle}. The DoF regions of these channels can be achieved via a combination of transmit and receive zero-forcing beamforming and time sharing. However, such schemes are sensitive to the imperfections in CSI at the transmitters (CSIT). It is known that in the complete absence of CSIT and when the transmit directions to the different receive antennas of a network are statistically indistinguishable, the DoF regions of MIMO broadcast and interference channels can be achieved with just receive zero-forcing beamforming and time sharing \cite{Chiachi2, Vaze_Dof_final, Zhu_Guo_noCSIT_DoF_2010, Vaze_Varanasi_Interf_Loclzn_2011}. 

With this background, it is of interest to investigate if there are channel models where interference alignment, and hence higher DoF compared with receiver zero-forcing and time-sharing schemes, can be achieved even under isotropic fading and without CSIT. To this end, it has been shown that in a compound channel setting, where channel matrices can take on one of finitely many values, it is possible to obtain interference alignment schemes even without CSIT to achieve higher DoF than attainable with a combination of just time sharing and receive zero-forcing \cite{GouJafarWang, W-S-Kramer, maddah-ali-compound-BC}. Subsequently, it was shown by the authors of this paper that cognition, where a transmitter is assumed to also know the other transmitters's message \cite{Devroye}, helps in expanding the DoF region, even in the absence of CSIT \cite{Vaze_Dof_final, Vaze_Dof_Cognitive_IC_ISIT}. Along a different direction, \cite{Jafar_correlations} studies a staggered block fading model, where the fading blocks corresponding to different users are assumed to be suitably misaligned and the transmitter(s) know the fading block boundaries but not the actual channel realizations. It is shown therein that by exploiting the suitable misalignment of the fading blocks, `blind' interference alignment can be effected at the receivers to improve the DoF. It was later proved that interference alignment is feasible when the channel matrices vary deterministically within a coherence block, instead of remaining constant (which is the usual assumption) \cite{Vaze_Dof_Cognitive_IC_ISIT}. Meanwhile, \cite{Jafar_Gou_blind_IA_2010} studied the MIMO BC with the receivers having reconfigurable antennas capable of switching between different preset modes. It is shown there that using reconfigurable antennas, the staggered block fading model can be created to achieve a DoF improvement. Subsequently, a similar conclusion is derived in \cite{lei_wang_mode_switching_IC} for the MIMO IC with the transmitters equipped with reconfigurable antennas. More recently, the authors of this paper proved that over the $2 \times 2 \times 2$ interference network, which is a layered, two-hop IC wherein the transmitters can communicate with the receivers only through an intermediate layer of relays, retro-cooperative interference alignment schemes can be developed even without CSIT, as long as there is a sufficient amount of (strictly delayed) feedback to the relays from the receivers \cite{vaze_varanasi_2hop_IC_allerton}.

\begin{figure}[!t] \centering
\includegraphics[scale=0.375]{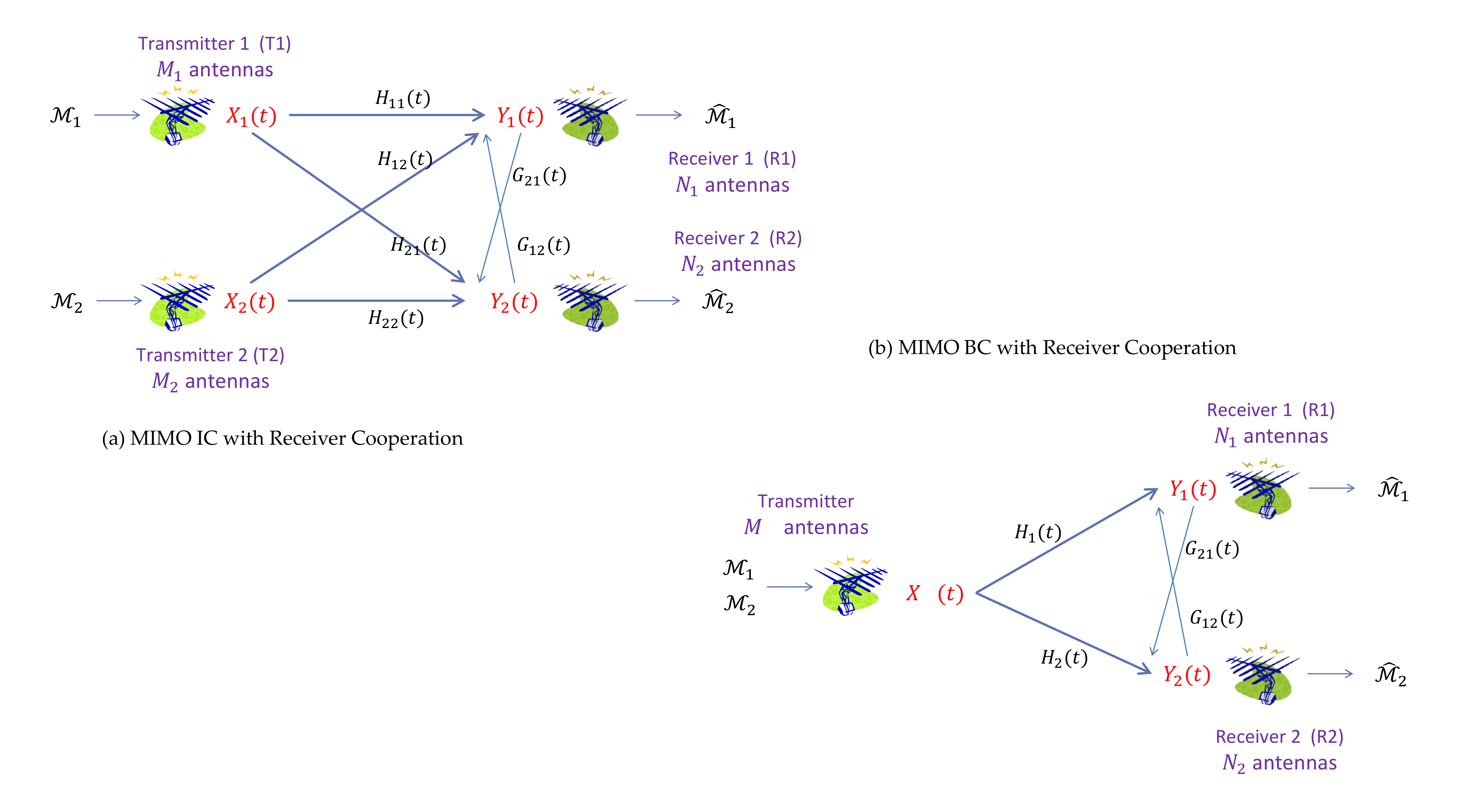}
\caption{The MIMO IC and the MIMO BC with no CSIT and with receiver cooperation}
\label{fig: channel model rx coop design fb}
\end{figure}

In this paper, we further explore this line of investigation by first noting that if the receivers are co-located, in which case the BC and the IC respectively reduce to the point-to-point and multiple access channels, the DoF remain unaffected by the presence or absence of CSIT \cite{Telatar}. This result motivates us to explore a model of receiver cooperation where the receivers have full-duplex radios so that they can also transmit signals over the same shared medium (see Fig. \ref{fig: channel model rx coop design fb}). Evidently, such a full-duplex receiver cooperation model in a BC or an IC results in a network that lies somewhere in between the two extremes of a BC or an IC without CSIT, respectively, on the one hand, and their co-located receivers (or idealized cooperation) counterparts on the other, namely, the point-to-point channel or the multiplex-access channel without CSIT, respectively. The full-duplex feature of the receivers naturally introduces the possibility of cooperation between receivers since the transmit signal of a receiver is heard by the other receiver. While such a receiver-cooperation model has been studied before with perfect and instantaneous CSIT \cite{prabhakaran_dest_coop_2011, Chiachi-Jafar} -- where it does not result in an enhancement of the DoF relative to the case of half-duplex receivers -- it is considered for the first time in this work in the more interesting setting of no CSIT and isotropic fading. In particular, we show that for the MIMO BC and the MIMO IC that this form of receiver cooperation can enhance the DoF even when the transmitters have no side-information whatsoever, thereby demonstrating for the first time that receiver cooperation can result in a DoF improvement. As a case in point, for the BC with a two-antenna transmitter, two single-antenna receivers, and no CSIT, the sum-DoF increase from $1$ for half-duplex receivers to $\frac{4}{3}$ for full-duplex receivers (see Theorem \ref{thm: dof region bc rx coop}). The DoF benefits associated with receiver cooperation are realized here using {\em retro-cooperative} interference alignment schemes. In these schemes, the receivers exchange information over the cooperative links that exist between them, and this information is designed such that a receiver can communicate useful information to the other receiver without creating any additional interference at it. Significantly, these schemes are shown to be DoF-region optimal for certain classes MIMO ICs and BCs, defined by certain relationships on numbers of antennas at different terminals (see Theorems \ref{thm: rx coop 2 2 1 1 generalize}, \ref{thm: rx coop 4 1 3 2 generalize}, and \ref{thm: dof region bc rx coop}). The converse results are proved through the DoF analysis of certain enhanced (genie-aided) networks and by showing that the resulting DoF upper bounds are indeed achievable.

The rest of the paper is organized as follows. Section \ref{sec:intro} introduces the two MIMO networks with no CSIT and with receiver cooperation that are the focus of our investigation. Section \ref{sec:mainresults} states all the results of this paper for the MIMO BC and MIMO IC. The proofs of the DoF outer bounds are provided in Section \ref{sec:outer} for the two MIMO networks with an arbitrary number of antennas at each terminal. Achievable schemes that establish DoF regions for certain classes of MIMO BCs and ICs are given in Sections \ref{sec:dof1} and \ref{sec:dof2}. Section \ref{sec:conclusions} concludes the paper.

\section{Models of Broadcast and Interference Channels with Receiver Cooperation but with no CSIT}
\label{sec:intro}

We introduce the models for broadcast and interference channels with receiver cooperation, where the receivers have full-duplex radios so that they can transmit signals over the same shared medium. To understand the impact of just receiver cooperation, we consider a scenario where the transmitters have no side information whatsoever. Moreover, in keeping with the no CSIT assumption, we assume that each receiver has no knowledge of its outgoing cooperation link, i.e., the outgoing channel to the other receiver. 

\subsection{The MIMO IC with Receiver Cooperation}
\label{subsec:modelic}

The input-output relationship for the $(M_1,M_2,N_1,N_2)$ MIMO IC with receiver cooperation is given by
\begin{eqnarray}
Y_1(t) & = & H_{11}(t) X_1(t) + H_{12}(t) X_2(t) + G_{12}(t) X_{R2}(t) + W_1(t) ~~\mbox{ and}\\
Y_2(t) & = & H_{21}(t) X_1(t) + H_{22}(t) X_2(t) + G_{21}(t) X_{R1}(t) + W_2(t), \label{eq: io relation rx coop}
\end{eqnarray}
where, at time $t$, $X_i(t) \in \mathbb{C}^{M_i \times 1}$ and $X_{Ri}(t) \in \mathbb{C}^{N_i \times 1}$ are the transmit signals of the $i^{th}$ transmitter and the $i^{th}$ receiver, respectively; $Y_i(t) \in \mathbb{C}^{N_i \times 1}$ is the received signal of the $i^{th}$ receiver; $W_i(t)$ is the additive noise at the $i^{th}$ receiver; $H_{ij}(t)$ and $G_{ij}(t)$ represent the channel matrices from the $j^{th}$ transmitter and from the $j^{th}$ receiver to the $i^{th}$ receiver, respectively; and there is a power constraint of $P$ at all terminals.

We consider here the case of Rayleigh fading and AWGN (additive white Gaussian noise). More precisely, the elements of channel matrices and additive noises are taken to independent and identically distributed (i.i.d.) according the zero-mean, unit-variance, circularly-symmetric complex Gaussian distribution (denoted henceforth as $\mathcal{C}\mathcal{N}(0,1)$). Moreover, their realizations are i.i.d. across time.

In order to be consistent with the assumption of transmitters having no side-information, it is assumed that each receiver has perfect knowledge of all channel matrices, except that of the one cooperative link that originates from it to the other receiver. Moreover, CSI at the receivers is taken to be instantaneous, without loss of generality. Specifically, the $i^{th}$ receiver knows channel matrices $\left\{ H_{ij}(t) \right\}_{i,j=1}^2$ and $G_{ij}(t)$ perfectly and instantaneously. Further, the radios at the receivers, though full-duplex, are not instantaneous, i.e., the transmit signal of a receiver can be a function of past channel matrices known to it and its past received signal. It cannot however depend on the current channel matrices and its current received signal.

Following \cite{Chiachi-Jafar}, we also define the idealized model of the MIMO IC with CSI and (receiver and transmitter) cooperation, wherein all terminals are assumed to have full-duplex radios as well as perfect and instantaneous CSI. In this model, the transmit signal of any given terminal can depend on the past and present channel states as well as on its past received signals. Thus, the input-output relationship for the MIMO IC with CSI and cooperation is analogous to the one in \eqref{eq: io relation rx coop}, except that the transmitters can also receive. It was shown in \cite{Chiachi-Jafar} however that full-duplex cooperation does not enhance the DoF region of the MIMO IC beyond the DoF region of its counterpart with half-duplex terminals with perfect and instantaneous CSI.

We define a few more settings for the sake of comparison. In all of these settings, the receivers are assumed to know all channel matrices perfectly, but they do not have full duplex radios (so that the input-output relationship is identical to that in \eqref{eq: io relation rx coop}, except that $X_{R1}(t) =  X_{R2}(t) = 0$):
\begin{itemize}
\item no CSIT: no side information at the transmitters;
\item perfect and instantaneous CSIT; 
\item delayed CSIT: transmitters know perfectly all the past channel matrices, but have no knowledge of the current channel matrices.
\item Shannon feedback: the transmitters know all past channel states as well as all past channel outputs.
\end{itemize}

The DoF regions for all the above models are defined in a standard manner.

The DoF regions of the IC with no CSIT \cite{Chiachi2, Vaze_Dof_final, Zhu_Guo_noCSIT_DoF_2010, Vaze_Varanasi_Interf_Loclzn_2011}, delayed CSIT \cite{Vaze_Varanasi_delay_MIMO_IC}, Shannon feedback \cite{Vaze_Varanasi_Shannon_fb_2user_IC_journal}, perfect and instantaneous CSIT \cite{Jafar-Maralle}, and CSI and cooperation \cite{Chiachi-Jafar} are known from the literature. If we denote the DoF regions under these five settings by $\mathbf{D}^{\rm no}$, $\mathbf{D}^{\rm d}$, $\mathbf{D}^{\rm S}$, $\mathbf{D}^{\rm p\& i}$, and $\mathbf{D}^{\rm coop}$, respectively, then in general we have
\[
\mathbf{D}^{\rm no} \subseteq \mathbf{D}^{\rm d} \subseteq \mathbf{D}^{\rm S} \subseteq \mathbf{D}^{\rm p\& i} = \mathbf{D}^{\rm coop}
\]
with the subset inclusions indicated above being strict for most of the MIMO ICs (for the last subset inclusion, please see \cite{Vaze_Varanasi_Shannon_fb_2user_IC_journal}).

\subsection{The MIMO BC with Receiver Cooperation}

The signal received by the receivers of the $(M,N_1,N_2)$ MIMO BC with receiver cooperation, where the transmitter has $M$ antennas and the receivers have $N_1$ and $N_2$ antennas, respectively, are given by
\begin{eqnarray}
Y_1(t) & = & H_1(t) X(t) + G_{12}(t) X_{R2}(t) + W_1(t) \mbox{ and} \\
Y_2(t) & = & H_2(t) X(t) + G_{21}(t) X_{R1}(t) + W_2(t).
\end{eqnarray}
Given the full description of the model for the MIMO IC, the details for the MIMO BC above should be evident. We omit them in order to avoid repetition except to restate that we consider the case of i.i.d. Rayleigh fading and AWGN and the transmitter has no side-information whatsoever; each receiver knows all channel matrices instantaneously, except the one that originates from it (whose realizations are not known it); and the radios at the receiver are full-duplex but not instantaneous -- the transmit signal of a receiver depends on the past channel matrices that are known to it and its past received signals but it can not depend on the current channel states or channel outputs.

The settings of no CSIT, delayed CSIT, Shannon feedback, and instantaneous CSIT are defined in a manner analogous to their IC counterparts in Section \ref{subsec:modelic}, and none of these settings involve receivers with full-duplex radios.
Their DoF regions are again defined in the standard manner. 

\section{Main Results}
\label{sec:mainresults}

This paper shows that retro-cooperative interference alignment schemes \cite{vaze_varanasi_2x2x2_allerton11} can be obtained using full-duplex receiver cooperation and that they yield a DoF benefit. These points are proved en route to characterizing the DoF regions of some important classes of MIMO BCs and ICs defined by certain relationships on numbers of antennas at the four terminals. Toward this end, we first derive outer-bounds to the DoF regions of the MIMO IC and BC with receiver cooperation. These bounds are later shown to be tight for certain classes of those channels.
\subsection{Outer-Bounds}
\label{sec:outerbounds}
The first two theorems below provide outer-bounds to the DoF region of the MIMO IC with receiver cooperation while the following two theorems specify outer bounds for the MIMO BC with receiver cooperation. 

\begin{theorem}[A BC-type Outer-bound for the IC] \label{thm: BCtype for IC rx coop}
The DoF region of the $(M_1,M_2,N_1,N_2)$ MIMO IC with receiver cooperation is outer-bounded by that of the $(N_1+N_2+M_1+M_2,N_1,N_2)$ MIMO BC with Shannon feedback. In particular, the former is outer-bounded by the following two inequalities:
\begin{eqnarray*}
\frac{d_1}{N_1+N_2} + \frac{d_2}{N_2} \leq 1 \quad \mbox{ and } \quad \frac{d_1}{N_1} + \frac{d_2}{N_1+N_2} \leq 1.
\end{eqnarray*}
\end{theorem}
\begin{IEEEproof}
See Section \ref{proof of outer-bounds: BCtype for IC rx coop}.
\end{IEEEproof}

\begin{theorem}[An IC-type Outer-bound for the IC] \label{thm: ICtype for IC rx coop}
The DoF region of the $(M_1,M_2,N_1,N_2)$ MIMO IC with receiver cooperation is outer-bounded by that of the $(M_1+k,M_2+N_1+N_2-k,N_1,N_2)$ MIMO IC with Shannon feedback, where $k$ is any non-negative integer less than or equal to $N_1+N_2$. Moreover, the former is outer-bounded by the following inequalities:
\begin{eqnarray*}
\frac{d_1}{\min(N_1+N_2,M_1)} + \frac{d_2}{\min(N_2,M_1)}  & \leq & \frac{N_2}{\min(N_2,M_1)} \mbox{ and }  \\
\frac{d_1}{\min(N_1,M_2)} + \frac{d_2}{\min(N_1+N_2,M_2)} & \leq & \frac{N_1}{\min(N_1,M_2)}.
\end{eqnarray*}
\end{theorem}
\begin{IEEEproof}
See Section \ref{proof of outer-bounds: ICtype for IC rx coop}.
\end{IEEEproof}

\begin{theorem}[A BC-type Outer-bound for the BC] \label{thm: BCtype for BC rx coop}
The DoF region of the $(M,N_1,N_2)$ MIMO BC with receiver cooperation is outer-bounded by that of the $(N_1+N_2+M,N_1,N_2)$ MIMO BC with Shannon feedback.
\end{theorem}
\begin{IEEEproof}
Follows from techniques developed in the proof of Theorem \ref{thm: BCtype for IC rx coop}.
\end{IEEEproof}

To derive the next bound for the BC with receiver cooperation, we need to invoke the model for the cognitive radio channel (CRC) \cite{Devroye}. In particular, we introduce the notation that the $(M_1,M_2,N_1,N_2,i)$ MIMO CRC with Shannon feedback is the $(M_1,M_2,N_1,N_2)$ MIMO IC with Shannon feedback, where in addition, the $i^{th}$ transmitter knows both messages (i.e., the $i^{th}$ transmitter is cognitive).

\begin{theorem}[A CRC-type Outer-bound for the BC] \label{thm: CRCtype for BC rx coop}
The DoF region of the $(M,N_1,N_2)$ MIMO BC with receiver cooperation is outer-bounded by that of the $(M,N_1+N_2,N_1,N_2,1)$ MIMO CRC with Shannon feeedback. In particular, the former is outer-bounded by the following two inequalities:
\begin{eqnarray*}
\frac{d_1}{\min(M,N_1+N_2)} + \frac{d_2}{\min(M,N_2)} & \leq & \frac{N_2}{\min(N_2,M)} \mbox{ and}\\
\frac{d_1}{\min(M,N_1)} + \frac{d_2}{\min(M,N_1+N_2)} & \leq & \frac{N_1}{\min(N_1,M)}.
\end{eqnarray*}
\end{theorem}
\begin{IEEEproof}
See Section \ref{proof of outer-bounds: CRCtype for BC rx coop}.
\end{IEEEproof}

In fact, all above outer-bounds are applicable even to a stronger setting. Specifically, all theorems hold for ICs and BCs with receiver cooperation as well as Shannon feedback.

\subsection{DoF Regions with Receiver Cooperation}

We characterize the DoF regions of certain classes of BCs and ICs with receiver cooperation. We start below with ICs where the receivers have an equal number of antennas.

\begin{theorem} \label{thm: rx coop 2 2 1 1 generalize}
For the MIMO IC with $N_1 = N_2 = N$, the DoF region with receiver cooperation is equal to $\mathbf{D}^{\rm coop}$ if $M_1 \leq N$ and/or $M_2 \leq N$; otherwise, it is given by
\begin{eqnarray*}
\mathbf{D}^{\rm rx-coop} = \left\{ (d_1,d_2) \left| d_1, d_2 \geq 0; ~ \frac{d_1}{\min (M_1,2N)} + \frac{d_2}{N} \leq 1; ~\frac{d_1}{N} + \frac{d_2}{\min (M_2,2N)} \leq 1 ~ \right. \right\}.
\end{eqnarray*}
\end{theorem}
\begin{IEEEproof}
When $M_1 \leq N_1=N_2=N$ and/or $M_2 \leq N_1=N_2=N$, the DoF region with CSI and cooperation can be attained without CSIT and without receiver cooperation \cite{Vaze_Dof_final}. Hence, it is sufficient to focus on the case, where $M_1,M_2>N$. The converse argument follows from Theorem \ref{thm: ICtype for IC rx coop}. What remains is the proof of  the achievability part for which new retro-cooperative interference alignment scheme is developed. The main idea of the scheme is illustrated first in Section \ref{sec: proof of thm: rx coop 2 2 1 1} through an example of the $(2,2,1,1)$ MIMO IC with receiver cooperation. The general case is proved in Section \ref{sec: proof of thm: rx coop 2 2 1 1 generalize}.
\end{IEEEproof}

We note that when the receivers have an equal number of antennas, the DoF region with receiver cooperation is equal to the DoF region with delayed CSIT, with the latter established in \cite{Vaze_Varanasi_delay_MIMO_IC}.

The DoF region of the MIMO IC is more elusive when $N_1 \neq N_2 $. Nevertheless, in what follows, we obtain the DoF region of a small class of MIMO ICs with an unequal number of antennas at the two receivers.

\begin{theorem} \label{thm: rx coop 4 1 3 2 generalize}
For the MIMO IC with $M_1 > N_1 > N_2 > \frac{N_2}{2} \geq M_2$ and $M_1+M_2 = N_1 + N_2$, the DoF region with receiver cooperation is given by
\[
\mathbf{D}^{\rm rx-coop} = \Big\{ (d_1,d_2) \Big| ~d_1, d_2 \geq 0; ~ d_2 \leq M_2; ~ d_1 + d_2 \leq N_1 \Big. \Big\},
\]
which is equal to the DoF region with CSI and cooperation.
\end{theorem}
\begin{IEEEproof}
The converse argument follows by noting that the DoF region with receiver cooperation can not be bigger than that with CSI and cooperation. We thus consider the achievability part, which is proved by developing a retro-cooperative interference alignment scheme. The main idea of this scheme is explained first through the example of the $(4,1,3,2)$ MIMO IC in Section \ref{sec: proof of thm: rx coop 4 1 3 2}. The general proof is given in Section \ref{sec: proof of thm: rx coop 4 1 3 2 generalize}.
\end{IEEEproof}

We next characterize the DoF region of the MIMO BC in which the receivers have an equal number of antennas.
\begin{theorem} \label{thm: dof region bc rx coop}
For the MIMO BC with $N_1 = N_2 = N$ and i.i.d. Rayleigh fading, the DoF region with receiver cooperation is given by
\[
\left\{ (d_1,d_2) \left| \frac{d_1}{\min(M,2N)} + \frac{d_2}{\min(M,N)} \leq \frac{N}{\min(M,N)}; ~ \frac{d_1}{\min(M,N)} + \frac{d_2}{\min(M,2N)} \leq \frac{N}{\min(M,N)} \right. \right\}.
\]
\end{theorem}
\begin{IEEEproof}
The converse follows from Theorem \ref{thm: CRCtype for BC rx coop}. On the achievability side, the case of $M < N$ is trivial. We thus consider the remaining case of $M > N$. It is sufficient to prove that the DoF pair
\[
Q \equiv \left( \frac{MN}{M+N}, \frac{MN}{M+N} \right)
\]
is achievable. Toward this end, note that the scheme developed in Section \ref{sec: proof of thm: rx coop 2 2 1 1 generalize} for achieving pair $P$ over the $(M_1,M_2,N,N)$ MIMO IC with receiver cooperation can be easily adapted to achieve point $Q$ over the $(M,N,N)$ MIMO BC. We omit details to avoid repetition.
\end{IEEEproof}

Note that the DoF region of the $(M,N,N)$ MIMO BC with receiver cooperation is equal to its DoF region with delayed CSIT found in \cite{Vaze-Varanasi-delay-MIMOBC}.

The main idea of how interference alignment is achieved with receiver cooperation is explained in the following remark.
\begin{remark}[On interference alignment with receiver cooperation] \label{rem: retro-coop ia rx coop}
During the initial phase, the receivers remain silent and listen to the transmitters for a sufficiently long duration of time so that they together have enough information required for decoding all data symbols sent by the transmitters. Next, the receivers exchange the signals they received over the previous phase. This exchange is accomplished such that a receiver can communicate new useful information to the other receiver without creating any additional interference at it, which manifests interference alignment.
\end{remark}

\section{Proofs of the Outer-Bounds}
\label{sec:outer}
In this section, we prove the outer-bounds stated in Section \ref{sec:outerbounds}.

\subsection{Proof of Theorem \ref{thm: BCtype for IC rx coop}}   \label{proof of outer-bounds: BCtype for IC rx coop}

It may be assumed without loss of generality that the $i^{th}$ receiver has $2N_i$ antennas of which $N_i$ are used strictly for reception and other $N_i$ are used strictly for transmission. Moreover, the signal transmitted by a receiver can be assumed to affect its own signal, and the receivers can be taken to know all channel matrices instantaneously. The idea of this outer-bound has been illustrated in Fig. \ref{fig: 2,2,1,1 IC rx coop converse} for the $(2,2,1,1)$ MIMO IC with receiver cooperation.

\begin{figure}[!h] \centering
\includegraphics[bb=0 235 1296 700,clip,scale=0.4]{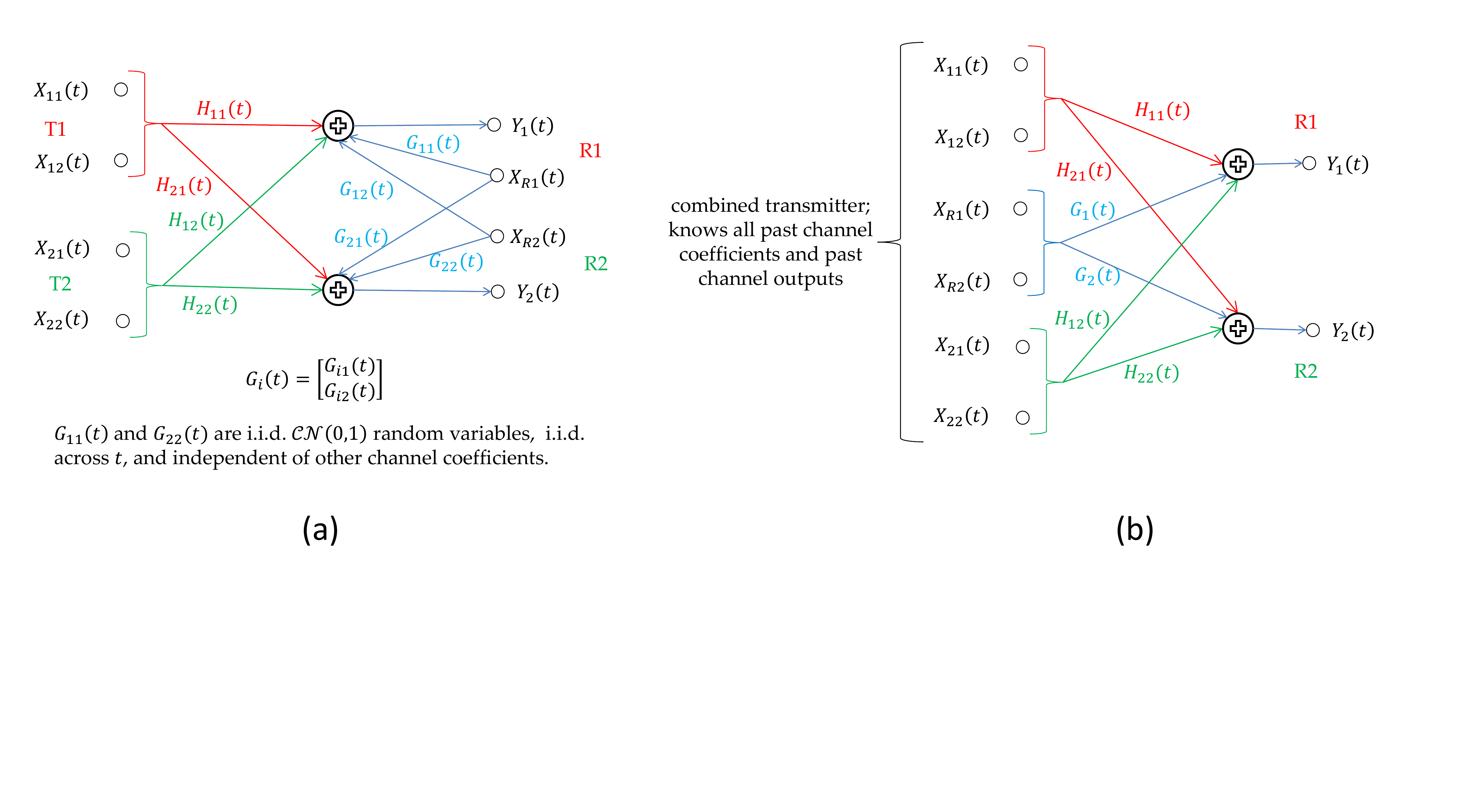}
\caption{A Converse Argument for the $(2,2,1,1)$-MIMO IC with Receiver Cooperation. The BC-type Outer-bound is tight for the $(2,2,1,1)$-MIMO IC with Receiver Cooperation.}
\label{fig: 2,2,1,1 IC rx coop converse}
\end{figure}

Suppose $G_{ii}(t)$ represents the channel matrix from the $i^{th}$ receiver to itself. Then, as per the above discussion, the input-output relationship of the given MIMO IC with receiver cooperation can be assumed, without loss of generality, to be as follows:
\[
Y_i(t) = H_{i1}(t) X_1(t) + H_{i2}(t) X_2(t) + G_{ii}(t) X_{Ri}(t) + G_{ij}(t) X_{Rj}(t) + W_i(t),
\]
where $j \not= i$, and the entries of $G_{11}(t)$ and $G_{22}(t)$ are i.i.d. $\mathcal{C}\mathcal{N}(0,1)$ random variables that are i.i.d. across time and independent of all other channel matrices and additive noises.

Consider now a $(M_1+M_2+N_1+N_2,N_1,N_2)$ MIMO BC with Shannon feedback, wherein the channel matrices are as follows:
\begin{eqnarray*}
H_1(t) & = & \begin{bmatrix} H_{11}(t) & H_{12}(t) & G_{11}(t) & G_{12}(t) \end{bmatrix} \mbox{ and }\\
H_2(t) & = & \begin{bmatrix} H_{21}(t) & H_{22}(t) & G_{21}(t) & G_{22}(t) \end{bmatrix}.
\end{eqnarray*}

We claim that any coding scheme feasible for the given MIMO IC with receiver cooperation is feasible for the above MIMO BC with Shannon feedback: This is because the transmitter of the above MIMO BC can construct its signal as
\[
X(t) = \begin{bmatrix} X_1(t) \\ X_2(t) \\ X_{R1}(t) \\ X_{R2}(t) \end{bmatrix},
\]
where $X_1(t)$, $X_2(t)$, $X_{R1}(t)$, and $X_{R2}(t)$ are the signals that T1, T2, R1, and R2 transmit at time $t$ over the given IC with receiver cooperation. Hence, any rate pair achievable over the given $(M_1,M_2,N_1,N_2)$ MIMO IC with receiver cooperation is also achievable over the $(M_1+M_2+N_1+N_2,N_1,N_2)$ MIMO IC with Shannon feedback, which implies that the DoF region of the former is outer-bounded by that of the latter.

\subsection{Proof of Theorem \ref{thm: ICtype for IC rx coop}}   \label{proof of outer-bounds: ICtype for IC rx coop}

The proof of this theorem is similar to that of the previous one. However, the result of Theorem \ref{thm: BCtype for IC rx coop} is strengthened here by not merging the two transmitters into a single one (through ideal transmitter cooperation). As argued in the proof of Theorem \ref{thm: BCtype for IC rx coop}, the rate pair achievable with receiver cooperation can be achieved in the absence of receiver cooperation provided we add $N_1+N_2$ extra antennas on the transmit side and give Shannon feedback to the transmitters. Furthermore, these extra antennas can be distributed in any manner between the two transmitters. In particular, $k$ of these extra antennas are given to one transmitter and the remaining to the other transmitter.

The last statement of the theorem is an application of the first one. The two inequalities stated correspond to the cases of $k = 0$ and $k = N_1+N_2$, respectively.


\subsection{Proof of Theorem \ref{thm: CRCtype for BC rx coop}}   \label{proof of outer-bounds: CRCtype for BC rx coop}

As before, the rate pair achievable with receiver cooperation can be achieved in the absence of it as long as we add an extra transmitter with $N_1+N_2$ antennas, give this new transmitter the message to be decoded by the second receiver, and give Shannon feedback to both transmitters. These enhancements of the channel result in the $(M_1,M_2,N_1,N_2,1)$ MIMO CRC with Shannon feedback and hence yield us the first statement of the theorem.

To derive the second part of the theorem, we need an outer-bound to the DoF region of the $(M_1,M_2,N_1,N_2,1)$ MIMO CRC with Shannon feedback. Toward this end, recall from \cite{Vaze_Varanasi_Shannon_fb_2user_IC_journal} that the inequality
\[
\frac{d_1}{\min(M_1,N_1+N_2)} + \frac{d_2}{\min(M_1,N_2)} \leq \frac{\min(N_2,M_1+M_2)}{\min(N_2,M_1)}
\]
has been shown to be an outer-bound to the DoF region of the $(M_1,M_2,N_1,N_1)$ MIMO IC with Shannon feedback. This proof given in \cite{Vaze_Varanasi_Shannon_fb_2user_IC_journal} is general enough in that it is applicable to the $(M_1,M_2,N_1,N_2,1)$ MIMO CRC with Shannon feedback. Hence, the inequality stated above is an outer-bound to the DoF region of the $(M_1,M_2,N_1,N_2,1)$ MIMO CRC with Shannon feedback, using which the first inequality stated in the theorem follows. The second inequality follows by symmetry.

\section{Proof of Theorem \ref{thm: rx coop 2 2 1 1 generalize}}
\label{sec:dof1}
In Section \ref{sec:ex1}, we illustrate the RCIA scheme through a simple example. The general case is proved in Section \ref{Sec:general1}.

\subsection{RCIA for the $(2,2,1,1)$ MIMO IC with Receiver Cooperation} \label{sec: proof of thm: rx coop 2 2 1 1}
\label{sec:ex1}
\begin{figure}[!h] \centering
\includegraphics[bb=0 250 1296 720,clip,scale=0.375]{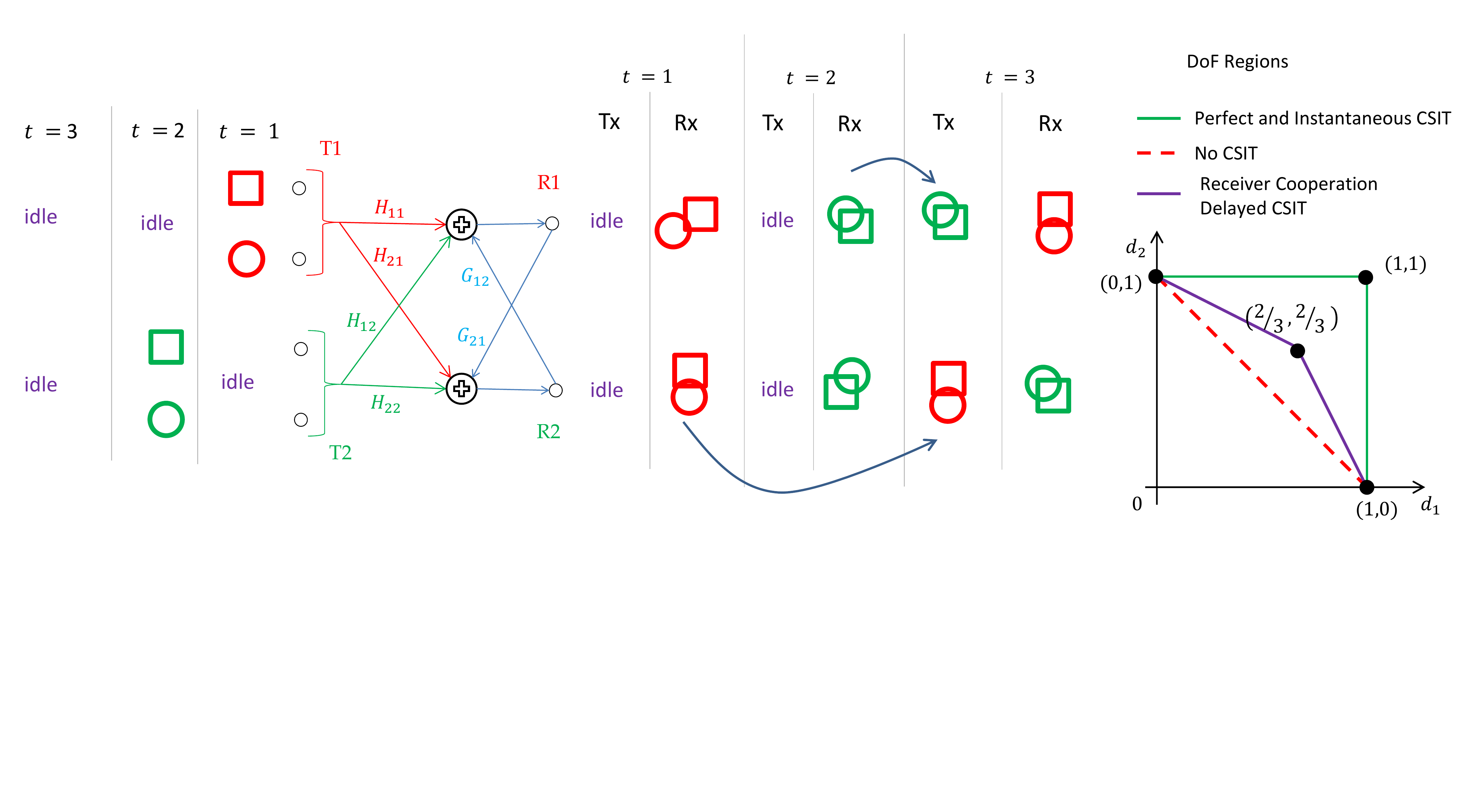}
\caption{A DoF-Region-Optimal Scheme for the $(2,2,1,1)$-MIMO IC with Receiver Cooperation}
\label{fig: 2,2,1,1 IC rx coop scheme and dof region}
\end{figure}

It is sufficient to prove that the pair $\left( \frac{2}{3},\frac{2}{3} \right)$ is attainable because if this pair can be achieved then the entire DoF region an be achieved via time sharing (see Fig. \ref{fig: 2,2,1,1 IC rx coop converse} for the shape of the DoF region). Toward this end, we will code over $3$ time slots and achieve $2$ DoF for each pair. The scheme works as follows; see also Fig. \ref{fig: 2,2,1,1 IC rx coop converse}.

At time $t = 1$, T1 transmits $2$ data symbols (DSs) intended for R1, while all other terminals do not send anything; see Fig. \ref{fig: 2,2,1,1 IC rx coop scheme and dof region}. Each receiver observes $1$ linear combination (LC) of these two DSs, and hence, R1 at this time can not decode two desired DSs\footnote{Throughout this paper, we ignore additive noises while dealing with the interference alignment schemes. This can be done because the presence of an additive noise can nt alter a DoF result.}. However, since the Rayleigh faded channel matrices are full rank almost surely, if R1 knows the signal received by R2 at $t = 1$, then it can successfully decode its symbols.

The scheme works analogously at time $t = 2$. T2 sends $2$ DSs intended for R2 while all other terminals do not transmit anything. Moreover, if the signal received by R1 at $t=2$ is revealed to R2, then R2 can decode the two desired DSs.

The swapping of the received signals is accomplished at $t = 3$, and this step achieves interference alignment. At this time, the transmitters remain silent, while R1 and R2 transmit the signals received by them at time $t = 2$ and $t = 1$, respectively; see Fig. \ref{fig: 2,2,1,1 IC rx coop converse}. Hence, R1 now knows the signal received by R2 at $t = 1$, which, as discussed before, would enable it to decode the two desired DSs. The operation of R2 is similar.

\begin{remark}[On interference alignment]
Here, interference alignment is achieved by noting that each receiver has seen only interference over one of the first two time slots, and hence they can exchange these interferences via the cooperation links between them. This allows each receiver to convey useful information to the other without causing any additional interference to it.

Since the DoF pair $\left( \frac{2}{3},\frac{2}{3} \right)$ is achievable with receiver cooperation as well as with delayed CSIT, it is instructive to compare the schemes that achieve this pair under these two models (see \cite{Vaze_Varanasi_delay_MIMO_IC} for the scheme with delayed CSIT). The operation of both schemes is identical over the first two time slots over which T1 and T2 transmit their DSs, and over the last time slot, both schemes involve swapping of the interferences seen earlier by the two receivers. In the case of delayed CSIT, this swapping is accomplished by noting that each transmitter can compute and thereby transmit the past interference it has created at its unpaired receiver. While such a technique is clearly infeasible over the present model, the same swapping of information can still be accomplished via receiver cooperation.
\end{remark}

\subsection{General Case}  \label{sec: proof of thm: rx coop 2 2 1 1 generalize}
\label{Sec:general1}


We observe from the shape of the DoF region that it is sufficient to prove the attainability of the DoF pair
\[
P \equiv \left( \frac{N \cdot M_1' \cdot (M_2'-N)}{N(M_2'-N)+M_2'(M_1'-N)}, \frac{N \cdot M_1' \cdot (M_2'-N)}{N(M_2'-N)+M_2'(M_1'-N)} \right)
\]
with $M_i' = \min(M_i,2N)$. The receiver-cooperation-based interference alignment scheme developed below to achieve pair $P$ is analogous to the delayed-CSIT-based scheme developed in \cite{Vaze_Varanasi_delay_MIMO_IC} to achieve the same pair $P$. The scheme consists of three phases and operates over $N(M_2'-N)+M_2'(M_1'-N)$ time slots.

Over the first phase, which consists of the initial $N(M_2'-N)$ time slots, T1 transmits $M_1'$ DSs per time slot intended for R1 while all other terminals remains silent. Since $M_1'>N$, R1 can not decode its signal. Consider a given time slot of this phase. If the signal received by R2 at its first $M_1'-N(\leq N)$ antennas is revealed to R1, R1 would be able to decode the DSs sent over this time slot (recall, Rayleigh-faded channel matrices are full-rank with probability $1$). Hence, in order to decode all desired DSs, R1 must learn a total of $N(M_2'-N)(M_1'-N)$ symbols that are available to R2.

The operation of the second phase is analogous. It takes $N(M_1'-N)$ time slots; T2 sends $M_2'$ DSs per time slot intended for R2; and R2 can decode DSs sent over a given time slot if it knows the signal received at that time by R1 at its first $M_2'-N(\leq N)$ antennas. Hence, over the last phase R2 needs $N(M_2'-N)(M_1'-N)$ symbols that are available to R1.

The last phase takes the remaining $(M_2'-N)(M_1'-N)$ time slot. Each receiver feeds symbols that are useful for the other at the rate of $N$ symbols per time slots. This allows each receiver to learn the missing set of linear combinations and thereby decode desired DSs.

\section{Proof of Theorem \ref{thm: rx coop 4 1 3 2 generalize}}
\label{sec:dof2}
We begin with an RCIA scheme for a specific example to give a simple illustration of the main idea in Section \ref{sec:ex2} and then present the general proof in Section \ref{sec:general2}.

\subsection{RCIA for the $(4,1,3,2)$ MIMO IC with Receiver Cooperation} \label{sec: proof of thm: rx coop 4 1 3 2}
\label{sec:ex2}
From the shape of the DoF region, we observe that it suffices to prove the achievability of DoF pair $(2,1)$. The achievability scheme  operates over $T = 2B+1$, $B>1$, time slots, and achieves the DoF pair $\left( \frac{4B}{2B+1},\frac{2B}{2B+1} \right)$. Hence, in the limit of $B \to \infty$, the scheme achieves the DoF pair $(2,1)$.

Total duration of $T$ time slots is divided into $B+1$ blocks, where each of the first $B$ blocks consist of $2$ time slots while the last block consists of just a single time slot. Hence, Block $b$, $b \leq B$, consists of time slots with index $2b-1$ and $2b$, whereas the $(B+1)^{th}$ block consists of the $(2B+1)^{th}$ time slot. The scheme has been presented in Fig. \ref{fig: 4,1,3,2 IC rx coop scheme and dof region} for $B = 2$.

Consider operation over Block $1$. At time $1$, T1 transmits DSs $\left\{ u_{i}^1 \right\}_{i=1}^4$ and T2 transmits DSs $v_1^1$, while the receivers do not send anything (superscript denotes the block index); see also Fig. \ref{fig: 4,1,3,2 IC rx coop scheme and dof region}. None of the receivers can do successful decoding at this time. To ensure that interference alignment can be achieved and decoding is eventually successful, we need to enable R1 to decode all DSs sent over $t = 1$. As we will see, this can be accomplished by communicating $Y_2(1)$ to R1. Hence, at time $t = 2$, transmit signals are constructed as follows: T2 sends $v_2^1$, R2 transmits $Y_2(1)$, while T1 and R1 remain silent. Hence, at $t = 2$, R1 knows $Y_1(1)$ and $Y_2(1)$. Since the Rayleigh faded channel matrices are invertible almost surely, R1, using $Y_1(1)$ and $Y_2(1)$, can evaluate all DSs sent by T1 and T2 at $t = 1$, namely, $\left\{u_i^1\right\}_{i=1}^4$ and $v_1^1$. On the other hand, R2 can decode $v_2^1$ at $t = 2$ (again via channel inversion), but not $v_1^1$. As we will see, R1 communicates $v_1^1$ to R2 at time $t = 4$ (i.e., over the second time slot of the next block).

With this motivation, consider Block $b = 2$; see also Fig. \ref{fig: 4,1,3,2 IC rx coop scheme and dof region}. The operation over this block is similar to that over Block $1$, except that at time $t = 4$ (i.e., over the second time slot of this block), R1 sends $v^1_1$. More precisely, at $t = 3$, T1 and T2 respectively transmit $\left\{ u_i^2 \right\}$ and $v_1^2$ while R1 and R2 remain silent. At $t = 4$, R1, T2, and R2 respectively transmit $v_1^1$, $v_2^2$, and $Y_2(3)$. As before, at $t = 4$, R1 can decode $\left\{ u_i^2 \right\}$ and $v_1^2$, whereas R2 can decode $v_2^2$ and $v_1^1$. Hence, at the end of Block $2$, all DSs sent to R2 over Block $1$ are successfully decoded.

\begin{figure}[!h] \centering
\includegraphics[bb=0 150 1296 720,clip,scale=0.4]{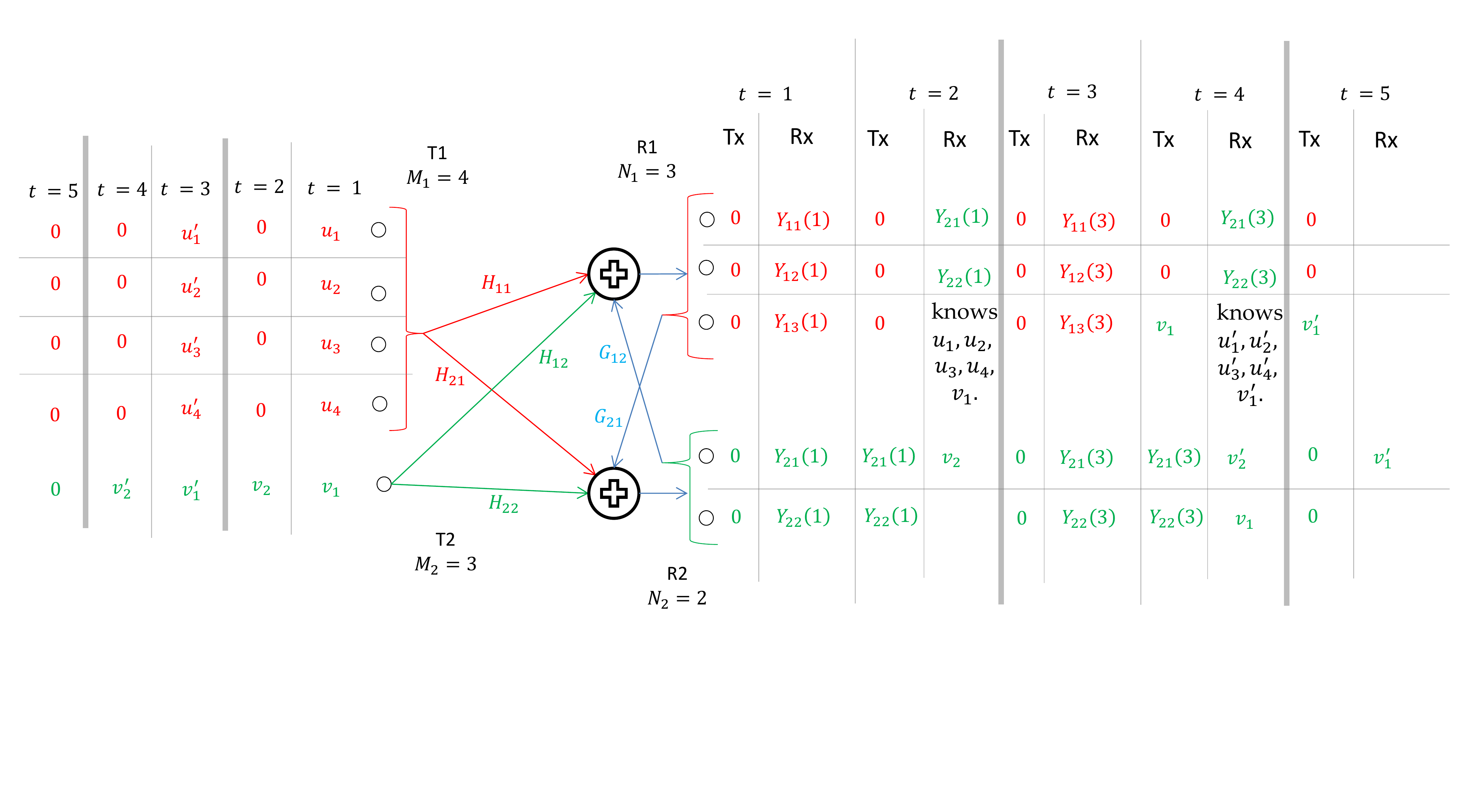}
\caption{A scheme achieving the DoF pair $(\frac{8}{5},\frac{4}{5})$ over the $(4,1,3,2)$-MIMO IC with receiver cooperation. This scheme in the limit of large number of blocks achieves the DoF Pair $(2,1)$, and hence, is asymptotically DoF-region optimal.}
\label{fig: 4,1,3,2 IC rx coop scheme and dof region}
\end{figure}

Now, the operation over Block $b$, $2 \leq b \leq B$, is analogous to that over Block $2$; we include details for the sake of completeness. At time $2b-1$, T1 and T2 transmit DSs $\left\{ u_i^b \right\}_{i=1}^4$ and $v_1^b$ respectively. By the previous discussion, R1 at the end of Block $b-1$ can decode $v_1^{b-1}$. Hence, at time $t = 2b$, R1, T2, and R2 can transmit $v_1^{b-1}$, $v_2^b$, and $Y_2(2b-1)$, respectively. As one would expect, R1 can decode DSs $\left\{ u_i^b \right\}_{i=1}^4$ and $v_1^b$, while R2 can decode $v_2^b$ and $v_1^{b-1}$. Hence, at the end of this block, R1 can decode all desired DSs sent until then, whereas R2 can decode all DSs sent over the first $b-1$ blocks and also $v_2^b$, which is sent over the second time slot of this block. Moreover, R1 knows DS $v_1^b$ at the end of this block.

Consider now the operation at $t = 2B+1$; see also Fig. \ref{fig: 4,1,3,2 IC rx coop scheme and dof region}. At $t = 2B$, R1 and R2 can decode all intended DSs sent to them over the first $B$ blocks, except for the decoding of $v_1^B$ at R2. The goal of the operation over this time slot is to ensure that $v_1^B$ is decodable at R2, and this can be accomplished because R1 knows this symbol at $t = 2B$. Thus, at time $t = 2B+1$, all terminals remain silent, except that R1 transmits $v_1^B$, which then becomes decodable at R2. Therefore, at $t = 2B+1$, decoding is successful at both receivers and the scheme concludes.

\begin{remark}[On interference alignment]
Interference alignment is achieved by first conveying to R1 the received signal of R2, thereby allowing R1 to decode all past data symbols sent by both transmitters. Later, R1 can transmit information that would ensure successful decoding at R2.
\end{remark}

\subsection{General Proof} \label{sec: proof of thm: rx coop 4 1 3 2 generalize}
\label{sec:general2}

It is sufficient to prove that the DoF pair $(N_1-M_2,M_2)$ is achievable. 

By coding over $T = 2B+1$ time slots, we achieve the DoF pair $\left( \frac{2(N_1-M_2)B}{2B+1},\frac{2M_2B}{2B+1} \right)$. Each block consists of two time slots, except for the last one. We describe the operation over Block $b$, $1 \leq b \leq B$. At time $t = 2b-1$, T1 and T2 send $M_1$ and $M_2$ DSs intended for their receivers, while the receivers remain silent. At time $t = 2b$, T2 transmits $M_2$ DSs, R2 sends the signal it has received at time $t = 2b-1$, while R1 transmits DSs sent by T2 at time $t = 2(b-1)-1$ (i.e., over the first time slot of the previous block). Note that over the second time slot of Block $1$, R1 remains silent. Moreover, in light of the discussion of the previous section, it is not difficult to see that at the end of Block $b$, $1 \leq b \leq B$, R1 can decode all DSs sent over that block. Hence, at time $t = 2b$, where $2 \leq b \leq B$, it can transmit DSs sent by T2 at $t = 2(b-1)-1$. Moreover, due to such an operation, R2 at time $t = 2b$, where $2 \leq b \leq B$, can decode all DSs sent over Block $b-1$. Finally, DSs sent to R2 at time $t = 2B-1$ can be made decodable at R2 by letting R1 transmit them at $t = 2B+1$.

\section{Conclusion}
\label{sec:conclusions}
The two-user MIMO BC and the MIMO IC with full-duplex receiver cooperation were investigated under the no CSIT assumption. By developing retro-cooperative interference alignment schemes, it is shown that receiver cooperation leads to a DoF improvement, even if the transmitters have no side information. The DoF regions are characterized completely for all MIMO BCs and MIMO ICs in which the numbers of antennas at the two receivers are equal. While our outer bounds are applicable for the general case with an arbitrary numbers of antennas at all the terminals for the MIMO BC and the MIMO IC, these bounds appear to be not tight in the general case where there are an unequal number of antennas at the two receivers. Nevertheless, in the case of the MIMO IC, the DoF region of a small class of channels with an unequal number of antennas at the two receivers is also established. The general case of unequal number of antennas at the two receivers is a topic for further investigation.

\bibliographystyle{IEEEtran}
\bibliography{references_DoF}

\begin{thebibliography}{10}
\providecommand{\url}[1]{#1}
\csname url@samestyle\endcsname
\providecommand{\newblock}{\relax}
\providecommand{\bibinfo}[2]{#2}
\providecommand{\BIBentrySTDinterwordspacing}{\spaceskip=0pt\relax}
\providecommand{\BIBentryALTinterwordstretchfactor}{4}
\providecommand{\BIBentryALTinterwordspacing}{\spaceskip=\fontdimen2\font plus
\BIBentryALTinterwordstretchfactor\fontdimen3\font minus
  \fontdimen4\font\relax}
\providecommand{\BIBforeignlanguage}[2]{{%
\expandafter\ifx\csname l@#1\endcsname\relax
\typeout{** WARNING: IEEEtran.bst: No hyphenation pattern has been}%
\typeout{** loaded for the language `#1'. Using the pattern for}%
\typeout{** the default language instead.}%
\else
\language=\csname l@#1\endcsname
\fi
#2}}
\providecommand{\BIBdecl}{\relax}
\BIBdecl

\bibitem{Shamai-W-S}
H.~Weingarten, Y.~Steinberg, and S.~Shamai, ``The capacity region of
  multiple-input multiple-output broadcast channels,'' \emph{IEEE Trans.
  Inform. Th.}, vol. 52, no. 9, pp. 3936--3964, Sep. 2006.

\bibitem{Chiachi-Jafar}
C.~Huang and S.~A. Jafar, ``Degrees of freedom of the {M}{I}{M}{O} interference
  \ channel with cooperation and cognition,'' \emph{IEEE Trans. Inform. Th.},
  vol. 55, no. 9, pp. 4211--4220, Sep. 2009.

\bibitem{Jafar-Maralle}
S.~A. Jafar and M.~J. Fakhereddin, ``Degrees of freedom for the {M}{I}{M}{O}
  interference channel,'' \emph{IEEE Trans. Inform. Th.}, vol. 53, no. 7, pp.
  2637--2642, Jul. 2007.

\bibitem{Chiachi2}
C.~Huang, S.~A. Jafar, S.~Shamai, and S.~Vishwanath, ``On degrees of freedom
  region of {M}{I}{M}{O} networks without channel state information at
  transmitters,'' \emph{IEEE Trans. Inform. Th.}, vol. 58, no. 2, pp. 849--857,
  Feb. 2012.

\bibitem{Vaze_Dof_final}
C.~S. Vaze and M.~K. Varanasi, ``The degree-of-freedom regions of {M}{I}{M}{O}
  broadcast, interference, and cognitive radio channels with no {C}{S}{I}{T},''
  \emph{IEEE Trans. Inform. Th.}, vol. 58, no. 8, pp. 5354 -- 5374, Aug. 2012.

\bibitem{Zhu_Guo_noCSIT_DoF_2010}
Y.~Zhu and D.~Guo, ``The degrees of freedom of {M}{I}{M}{O} interference
  channels without state information at transmitters,'' \emph{IEEE Trans.
  Inform. Th.}, vol. 58, no. 1, pp. 341--352, Jan. 2012.

\bibitem{Vaze_Varanasi_Interf_Loclzn_2011}
C.~S. Vaze and M.~K. Varanasi, ``A new outer-bound via interference
  localization and the degrees of freedom regions of {M}{I}{M}{O} interference
  networks with no {C}{S}{I}{T},'' \emph{in print, IEEE. Trans. Inform. Th.},
  July 2012, Available: http://arxiv.org/abs/1105.6033.

\bibitem{GouJafarWang}
T.~Gou, S.~A. Jafar, and C.~Wang, ``On the degrees of freedom of finite state
  compound wireless networks,'' \emph{IEEE Trans. Inform. Th.}, vol. 57, No. 6,
  pp. 3286--3308, June 2011.

\bibitem{W-S-Kramer}
H.~Weingarten, S.~Shamai, and G.~Kramer, ``On the compound {M}{I}{M}{O}
  broadcast channel,'' in \emph{Inform. Th. and Applications Workshop}, UCSD,
  San Diego, USA, Jan. 2007.

\bibitem{maddah-ali-compound-BC}
M.~A. Maddah-Ali, ``On the degrees of freedom of the compound {M}{I}{M}{O}
  broadcast channels with finite states,'' Sep. 2009, Available:
  http://arxiv.org/abs/0909.5006v3.

\bibitem{Devroye}
N.~Devroye, P.~Mitran, and V.~Tarokh, ``Achievable rates in cognitive radio
  channels,'' \emph{IEEE Trans. Inform. Th.}, vol. 52, no. 5, pp. 1813--1827,
  May 2006.

\bibitem{Vaze_Dof_Cognitive_IC_ISIT}
C.~S. Vaze and M.~K. Varanasi, ``The degrees of freedom region of the
  {M}{I}{M}{O} cognitive interference channel with no {C}{S}{I}{T},'' in
  \emph{IEEE Intl. Symp. Inform. Th.}, Austin, USA, June 2010.

\bibitem{Jafar_correlations}
S.~A. Jafar, ``Blind interference alignment,'' \emph{IEEE Journal of Selected
  Topics in Signal Processing}, vol. 6, no. 3, pp. 216--227, June 2012.

\bibitem{Jafar_Gou_blind_IA_2010}
C.~Wang, T.~Gou, and S.~A. Jafar, ``Aiming perfectly in the dark - blind
  interference alignment through staggered antenna switching,'' \emph{IEEE
  Trans. Signal Processing}, vol. 59, no. 6, pp. 2734--2744, June 2011.

\bibitem{lei_wang_mode_switching_IC}
L.~Ke and Z.~Wang, ``Degrees of freedom regions of two-user {M}{I}{M}{O} {Z}
  and full interference channels: The benefit of reconfigurable antennas,''
  \emph{IEEE Trans. Inform. Th.}, vol. 58, no. 6, pp. 3766--3779, June 2012.

\bibitem{vaze_varanasi_2hop_IC_allerton}
C.~S. Vaze and M.~K. Varanasi, ``The degrees of freedom of the 2x2x2
  interference network with delayed {C}{S}{I}{T} and with limited shannon
  feedback,'' in \emph{49th Annual Allerton Conf. Comm., Control and
  Computing}, Monticello, IL, USA, Sep. 2011, submitted for review, July 2011
  (see arXiv:1109.5790)",.

\bibitem{Telatar}
I.~E. Telatar, ``Capacity of multi-antenna {G}aussian channels,'' \emph{Euro.
  Trans. Telecomm.}, vol. 10, no. 6, pp. 585--595, Nov./Dec. 1999.

\bibitem{prabhakaran_dest_coop_2011}
V.~M. Prabhakaran and P.~Viswanath, ``Interference channels with destination
  cooperation,'' \emph{IEEE Trans. Inform. Th.}, vol. 57, no. 1, pp. 187--209,
  Jan. 2011.

\bibitem{Vaze_Varanasi_delay_MIMO_IC}
C.~S. Vaze and M.~K. Varanasi, ``The degrees of freedom region and interference
  alignment for the {M}{I}{M}{O} interference channel with delayed
  {C}{S}{I}{T},'' \emph{IEEE Trans. Inform. Th.}, vol. 58, no. 7, pp.
  4396--4418, July 2012.

\bibitem{Vaze_Varanasi_Shannon_fb_2user_IC_journal}
------, ``The degrees of freedom region of the {M}{I}{M}{O} interference
  channel with {S}hannon feedback,'' \emph{submitted, IEEE Trans. Inform. Th.},
  Oct. 2011, Available: http://arxiv.org/abs/1109.5779.

\bibitem{vaze_varanasi_2x2x2_allerton11}
------, ``The degrees of freedom of the 2x2x2 interference network with delayed
  {C}{S}{I}{T} and with limited shannon feedback,'' in \emph{49th Annual
  Allerton Conf. Comm., Control and Computing}, Monticello, IL, USA, Sep. 2011,
  submitted for review, July 2011 (see arXiv:1109.5790)",.

\bibitem{Vaze-Varanasi-delay-MIMOBC}
------, ``Degrees of freedom region for the two-user {M}{I}{M}{O} broadcast
  channel with delayed {C}{S}{I}{T},'' in \emph{IEEE Intl. Symp. Inform. Th.},
  Aug. 2011.

\end{thebibliography}
\end{document}